# A geometrical description of charge distribution on the disordered conductor surfaces

[1] **Mehdi Jafari Matehkolaee**, [2]**Narjes Sadri**

[1] *Department of Physics, Amirkabir University of Technology (Tehran Polytechnic) P.O.Box:15875-4413, Tehran, Iran*

,[2] *Department of Physics, Alzahra University, Tehran 19938-93973, Iran*

m.matehkolaee@aut.ac.ir

## Abstract

This paper considers the distribution of charge over the surface of a conductor, and specifically the old rule-of-thumb that charge accumulates near sharp points and more generally in regions of high curvature. The discussion is almost entirely qualitative. Various conductors with different geometric shapes have been considered and the charge density on their points has been compared. Our discussion shows that the statement "surface charge density is higher at points with greater curvature" does not seem to be true and at least it can be modified. We have avoided unnecessary relationships as much as possible and tried to follow the discussion qualitatively.

## I. Introduction

In most of the textbooks on the subject of electrostatics, the issue of surface charge density and generally charge distribution on the surfaces of conductors is discussed. However, this discussion is challenging and somewhat unclear for most students.

As usual, most students who face this problem are looking for solution to this question:

" Why do electric charges tend to go towards the sharp points of conductors? " In the pedagogical article [1], an attempt has been made to solve this question. Also, another pedagogical paper, is more focused on this question [2]. Perhaps the attention to the point raised in this article is convincing and a proper justification for the mentioned question was raised. That is, the potential of internal points of the non-conducting object is always greater than the exterior points. Therefore, after excitation of this object, electric charges will tend (based on the intrinsic translation of charges from higher to lower potential) to reach the outer points, edges and corners. For example, in the following figure, a three-dimensional non-conducting object, the potential of point D is always higher than other points.

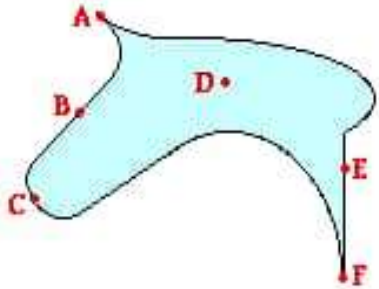

Figure (1): A closed three-dimensional non- conducting object.

It is clear that the quantity of surface charge density depends on the geometry of the conductors. However, in general, there is no unique relationship between curvature and this quantity [3]. Based on the report [4] the charge density is directly proportional to the fourth root of the Gaussian curvature of the surface. In this case, there is also a more detailed and exact survey [5]. In [6] has reviewed the previous experiments and attempts and tried to obtain an analytical formula for relating charge density with the surface curvature of conductors.

Apart from all these attempts mentioned in the scientific literature, there is a classic exercise in the standard book [7] that confirms the relationship between the magnitude of the normal derivative of the electric field and the curvature of the conductive surface.

$$\frac{\partial E}{\partial n} = -E\left(\frac{1}{R_1} + \frac{1}{R_2}\right) \tag{1}$$

where $E$ is the electrostatic field, $R_1$ and $R_2$ are the two principal radii of curvature of the surface at the point in question. Note that, the subject of principal radii of curvature and the relation between area and radii of curvature are aspects of differential geometry which are perhaps not standard fare for most undergraduate students. Therefore, searching for differential geometry books will be useful for these students.

There is fascinating and alternate evidence for the equation (1), without using of Gauss' law, which is highly valuable and helpful as a geometrical procedure [8].

## II. A review discussion

As mentioned in the introduction (based on the references), many efforts have been made to show the dependence of the quantity of surface charge density on the curvature of closed conductor surfaces. Despite all this, with a simple example, it can be shown that this issue is somewhat valid or at least it is not considered a generally correct statement. We start the discussion here with a challenging example.

Suppose we fold a sheet of metal, according to the following figure, it is clear that its curvature is still zero but the charge density on the line of fold is large!

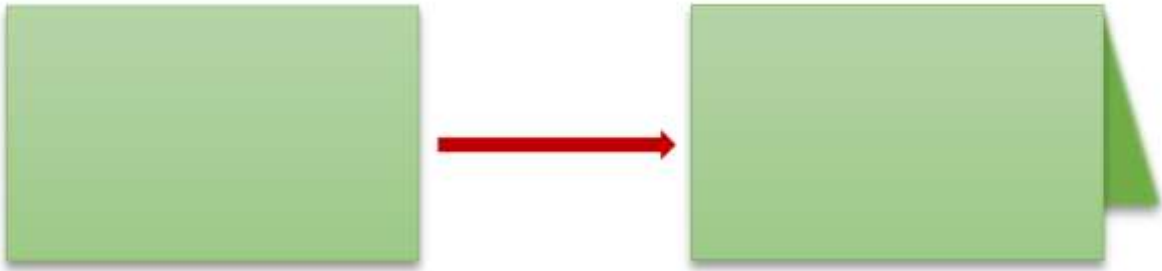

Figure (2): A conducting sheet with a certain charge density is folded in the right figure. It is quite obvious that the curvature is still zero in the folded sheet but the charge density at the folding line is large.

From figure (2), we can at least get the following result:

The surface charge density doesn't have straight-forward relation with the curvature of conductors.

The next example is the closed conductor below. By drawing the internal tangent spheres at points A, B and C, it is clear that the amount of curvature is the same at these three points.

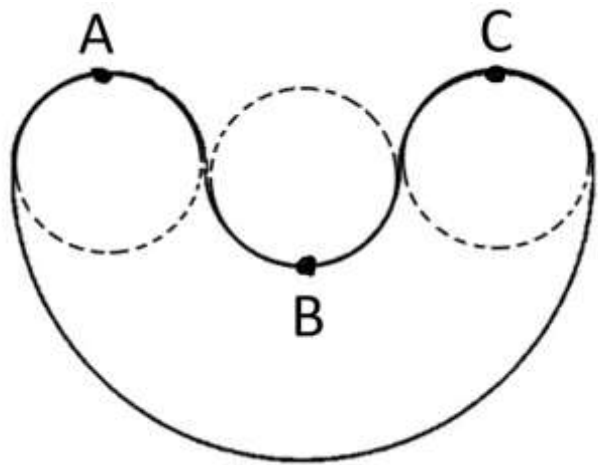

Figure (3): A three-dimensional closed conductor, the radii of the tangent spheres at three points A, B and C are the same.

We spray electric charge on this surface and stand aside and wait for the electric charges to make all its points the same potential (as an equipotential area) by moving on the conductive surface. Now is the time to ask this question. At which point is the surface charge density greater? When the curvature of the points is the same, what criterion should we consider comparing the charge density of the points?

To find the desired criterion, pay attention to the figure below

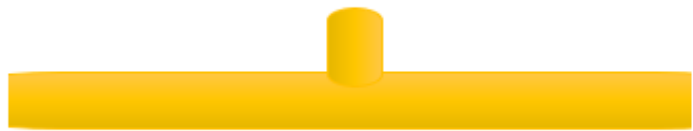

Figure (4): A three-dimensional closed conducting rod that has a pimple. The charge density on this surface is known.

In figure (4) we see a flat three-dimensional closed conductor with a pimple on it. Now the two ends of this rod-shaped conductor can be rotated in two ways. If we rotate upwards, we will have the following figure

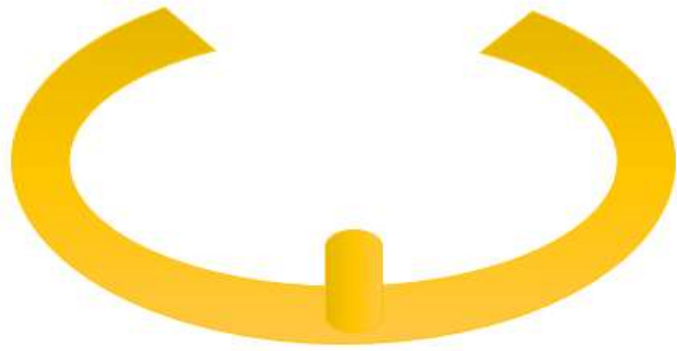

Figure (5): A three-dimensional conducting rod whose two ends have been turned upwards and brought near to each other.

By gradually bringing the two ends of this conductive rod together, we will have a closed conductor in which the electric field will tend to zero. In this way, the charge density on the pimple tends to be zero, but the curvature on it is still larger than in other places!

Now, if we turn the two ends of the rod in figure (4) downwards, we get a different result.

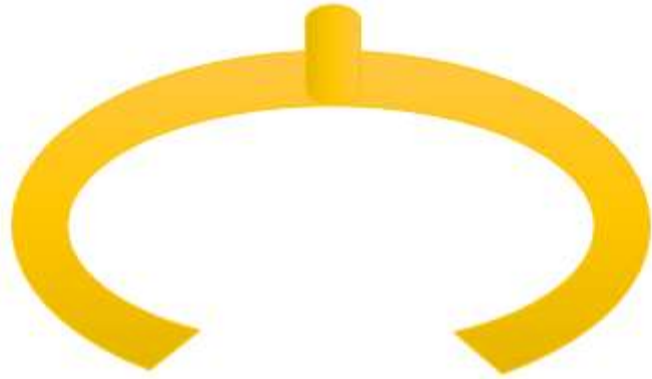

Figure (6): A three-dimensional conductive rod whose two ends have been turned downwards and brought near to each other.

It is quite clear that in this situation the charge density on the pimple is higher than in other points. Therefore, the criteria we have been looking for can be presented as follows:

" Under the condition of the same curvature, the electric charge density on the convex point is greater than the concave point of the surface of a conductor ".

So, in Figure (3), the charge density at points A and C is higher than at point B. This discussion can be extended to other situations as well.

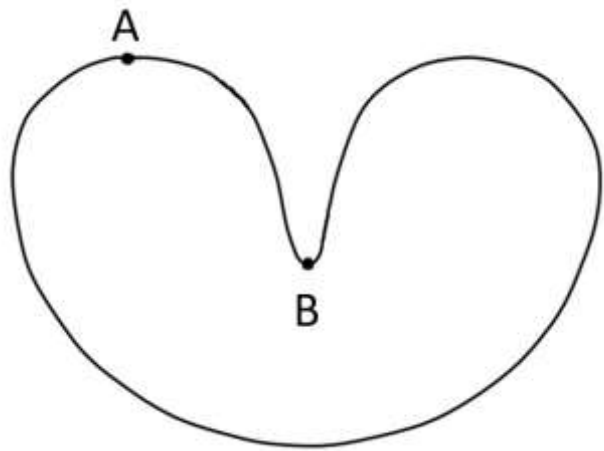

Figure (7): A closed conductor. Obviously, concerning the figure, the curvature at B is much higher than at A.

Although the curvature in A is much larger than B, the discussion dictates that the charge density in A is larger than B. Finally, the comparison of the density on the two points indicated in the figure below is clear and does not need to be discussed further.

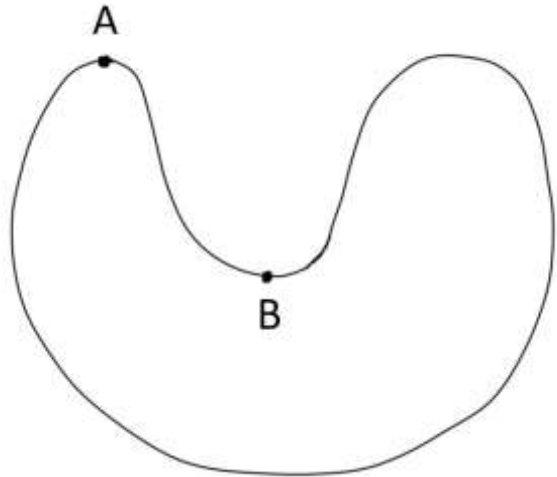

Figure (8): A closed conductor. In this figure, the curvature at A is higher than B and without hesitation, the charge density at A is larger than B.

From the discussion of this section, it can be deduced that, in general, the charge density in convex points of closed conductors is greater than in concave points.

## III. A good example: knotted torus

In general, as an electrostatic problem, the torus is not a simple form, and even finding the potential corresponding to a charge distributed on a torus is not easy [9].

In our opinion, an adequate example for the mentioned discussion is comparing charge density on the two points on the surface of a knotted torus.

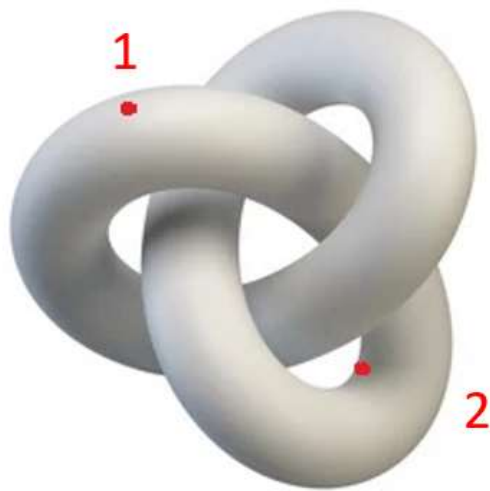

Figure (9): A knotted torus. The aim of this figure is to compare surface charge density between points 1 and 2.

At first, it is important to notice that at point 1 both curvature radii are positive. However, at point 2, one of them is positive and the other is negative. For a better understanding of this scenario, see the different signs of curvature at point 2:

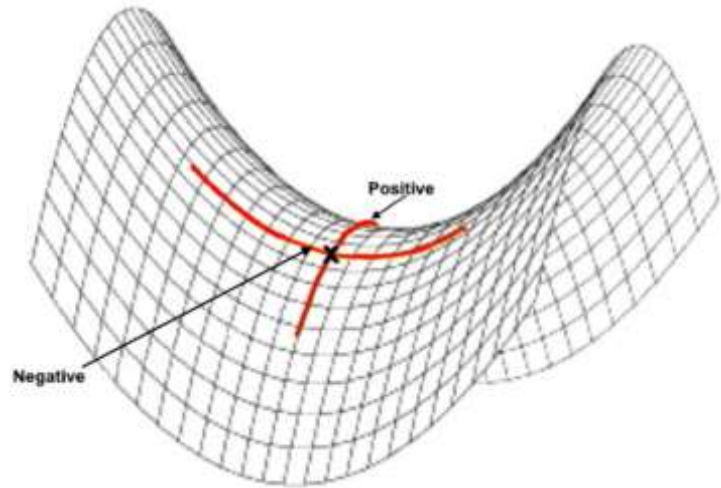

Figure (10): Different signs of curvature at point 2 related to figure (9)

From the figure, it is clear that for point 2, we have to know where the absolute value of the negative curvature radius is bigger or the positive one. So we can estimate from figure (9) that minus the right-hand side of equation (1) is bigger for point 1. Therefore, for point 1 the electric field grows faster as one approaches the surface.

If we want to make a comparison according to the discussion of the previous part, relying on the statement we concluded, the charge density at point 1 is higher than at point 2.

**IV. Conclusion**

The only used equation in this article is equation (1). However, this equation is not at all simple for undergraduate students. Most introductory university students, even those in an honors class, would not have yet seen partial derivatives such as used in equation (1). Nor would they have had a class in differential geometry. So the phase "fourth root of the Gaussian curvature of the surface" (referring to report of ref. [4]) would be meaningless to them. For these reasons, it seems necessary to read at least introductory differential geometry in order to follow the outline of this paper.

Perhaps one of the most interesting points related to the subject of the distribution of electric charges on the closed conductor's surface is that the charge density on concave points is always smaller than on convex points. In our opinion, this article can be useful and open the way for understanding this subject, which is relatively challenging for undergraduate students.

**Acknowledgement**

I thank Professor Mohammad Reza Sarkardei for the interesting discussions and comments, and Professor Mohammad Khorrami to the critical reading.

**References**

[1] H S. Fricker, *Why does charge concentrate on points?*, Phys.Educ.**24**, (1989).

[2] M. Jafari Matehkolaee and A.Naderi Asrami, *The review on the charge distribution on the conductor surface*, European J of Physics Education,4(3) (2013).

[3] I W McAllister, *Conductor curvature and surface charge density,* J.Phys.D:Appl.Phys.**23**,359-362(1990).

[4] Kun-Mu Lio, *Relation between charge density and curvature of surface of charged conductor*, Am.J.Phys.,**55**,9(1987).

[5] Daniel J. Cross, *When the charge on a planer conductor is a function of its curvature*, J.Math.Phys.**55**,123504 (2014).

[6] K.Bhattacharya, *On the dependence of charge density on surface curvature of an isolated conductor*, Phys.Scr.**91** 035501(2016).

[7] J.D.Jackson, *Classical Electrodynamics* (Wiley, New York,1999).

[8] Richard C Pappas, *Differential-geometric solution of a problem in electrostatics*, SIAM Review, 28(2):225–227, 1986.

[9] J. Lekner, *Electrostatics of a family of conducting toroids,* Eur. J. Phys. **30**, 477(2009).